\documentclass[12pt]{iopart}
\usepackage{graphicx}

\newcounter{parentequation}
\def\beglet{
  \addtocounter{equation}{1}%
  \setcounter{parentequation}{\value{equation}}%
  \setcounter{equation}{0}%
  \def\theequation{\arabic{parentequation}\alph{equation}}%
  \ignorespaces
}
\def\endlet{
  \setcounter{equation}{\value{parentequation}}%
  \def\theequation{\arabic{equation}}%
  \addtocounter{equation}{0}%
}
\def\ltsima{$\; \buildrel < \over \sim \;$}
\def\gtsima{$\; \buildrel > \over \sim \;$}
\def\simlt{\lower.5ex\hbox{\ltsima}}
\def\simgt{\lower.5ex\hbox{\gtsima}}

\begin{document}

\title[The Lyth Bound Revisited]{The Lyth Bound Revisited }

\author{George Efstathiou$^{1}$ and Katherine J. Mack$^{2}$
\footnote[3]{To
whom correspondence should be addressed (gpe@ast.cam.ac.uk)}
}

\address{[1]  Institute of Astronomy, Madingley Road, Cambridge,
CB3 OHA, UK}

\address{[2]
Department of Astrophysical Sciences, Princeton University,
Peyton Hall - Ivy Lane, Princeton, New Jersey 08544, USA}

\begin{abstract}
We investigate the Lyth relationship between the tensor-scalar ratio,
$r$, and the variation of the inflaton field, $\Delta \phi$, over the
course of inflation. For inflationary models that produce at least $55$ 
e-folds of inflation, there is a correlation between $r$ and
$\Delta \phi$ as anticipated by Lyth, but the scatter around the
relationship is huge.  However, for inflationary models that satisfy
current observational constraints on the scalar spectral index and its
first derivative, the Lyth relationship is much tighter. In particular,
any inflationary model with $r \simgt 10^{-3}$ must have $\Delta
\phi  \simgt m_{pl}$. Large field variations are therefore
required if a tensor mode signal is to be detected in any foreseeable
cosmic microwave background (CMB) polarization experiment.

\end{abstract}




\section{Introduction}

The inflationary paradigm, promoted by Guth (1981), Linde (1982) and
others, has achieved spectacular success in explaining the acoustic
peak structure seen in the CMB (see Bennett \etal 2003 and references
therein). Nevertheless, very little is known about the mechanism of
inflation and how it is related to fundamental physics. The simplest
mechanisms predict an adiabatic spectrum of nearly scale-invariant,
Gaussian, primordial fluctuations (for reviews see Linde 1990, Lyth
and Riotto 1999, Liddle and Lyth 2000). However, as Peiris \etal
(2003, hereafter P03) and Kinney \etal (2004) demonstrate, a wide class of
phenomenological inflationary models are compatible with the CMB and
other cosmological data.

Given our ignorance of the underlying physics, it is worth asking
what general statements can be made about inflation from particular
types of observation. For example, it is well known ({\it e.g.} Lyth 1984)
that the amplitude
of the tensor mode CMB anisotropy fixes the energy scale of
slow-roll inflation
\begin{equation}
 V^{1/4} \approx 3.3 \times 10^{16} r^{1/4}  \; {\rm GeV}, \label{I1}
\end{equation}
where $r$ is the relative amplitude of the tensor and scalar modes
defined as in P03.\footnote{$r = \Delta^2_h(k_0)/\Delta^2_{\cal R}(k_0)$,
where $\Delta^2_h$ and $\Delta^2_{\cal R}$ are the power spectra of
the tensor and scalar curvature perturbations defined at a fiducial
wavenumber of $k_0 = 0.002 \;{\rm  Mpc}^{-1}$.} The present $95\%$ upper
limit of $r < 0.36$ (Seljak \etal 2004) gives the constraint
$V^{1/4} \simlt 2.6 \times 10^{16}\; {\rm GeV}$, or equivalently
$V \simlt 2.2 \times 10^{-11} m^4_{\rm pl}$.

However, a successful model of inflation must also reproduce the
observed amplitude of scalar perturbations and produce a
sufficiently large number of e-foldings, $N(k)$, between the end of
inflation and the time that these perturbations were
generated. Constraints on more general inflation models can be
analysed using a set of `inflationary flow' equations
(Hoffman and Turner 2001, Kinney 2002, Easther and Kinney 2003, P03,
Kinney \etal 2004) describing the evolution of a hierarchy of
`slow-roll' parameters. The first two slow-roll parameters
are defined in terms of the Hubble parameter, $H$, during inflation by
\beglet
\begin{eqnarray}
\epsilon &=& {m^2_{\rm pl} \over 4 \pi}\left ( H^\prime \over H \right)^2, \\
\sigma &=& {m^2_{\rm pl} \over  \pi} \left ({1 \over 2} {H^{\prime\prime} \over H}
- \left( {H^\prime \over H } \right )^2 \right ), \label{I2}
\end{eqnarray}
\endlet
where primes denote differentiation of $H(\phi)$ with respect to
the inflaton  $\phi$. In terms of $\epsilon$, the derivatives of
$\phi$ and $H$ with respect to the number of e-foldings to the end
of inflation are given by
\beglet
\begin{equation}
{d \phi \over dN} =  {m_{\rm pl} \over 2 }\sqrt {\epsilon \over \pi}, \label{I3}
\end{equation}
\begin{equation}
{1 \over H} {d H \over dN} =  \epsilon, \label{I4}
\end{equation}
\endlet
and $\epsilon < 1$ is a  necessary criterion for inflation. To second
order in slow-roll parameters, the tensor-scalar ratio $r$ is given
by 
\begin{equation}
r = 16 \epsilon [ 1 - C(2 \epsilon + \sigma)], \label{I5}
\end{equation}
where $C= 0.08145$ (see Liddle, Parsons and Barrow 1994). 

Lyth (1997) noted that for slow-roll inflation, equations (\ref{I3}) and (\ref{I5}) 
can be used to relate the change  in the inflaton 
during inflation, $\Delta \phi$,  to the tensor-scalar ratio $r$, since if $\epsilon$
is roughly constant 
\begin{equation}
\Delta \phi \sim {m_{\rm pl} \over 8}  \left ( {r \over \pi} \right )^{1/2} 
\Delta N. \label{I6}
\end{equation}
Since the Universe inflates by $\Delta N \approx 4$ during the period
that  wavelengths corresponding to the CMB multipoles 
$2 \le \ell \le 100$ cross  the Hubble radius, equation (\ref{I6}) leads to a
bound between $\Delta \phi$ and $r$, 
\begin{equation}
\Delta \phi >  m_{\rm pl} (r /4\pi)^{1/2},      \label{I7}     
\end{equation}
which we will refer to as the `Lyth bound.' According to this bound,
high values of $r$ require changes in $\Delta \phi$ of order $m_{\rm
pl}$.  In fact, equation (\ref{I7}) is a rather crude bound, since at
least 50-60 e-foldings are required before inflation ends (see {\it
e.g.}  Liddle and Leach 2003). Over the full course of inflation
$\Delta \phi$ could therefore exceed (\ref{I7}) by an order of
magnitude or more.  

In this paper, we investigate the Lyth bound, making as few
assumptions as possible about the nature of inflation. Our aim is to
establish whether a relation of the form (\ref{I7}) applies to general
models of inflation that are compatible with observational constraints
on the scalar fluctuation spectrum.  Lyth (1997), Liddle and Lyth
(2000), Kinney (2003) and others argue that inflation cannot be
described by a low energy effective field theory if $\Delta \phi
\simgt m_{pl}$ and so high values of $r \sim 1$ are possible only in
models for which no rigorous theoretical framework exists. Conversely,
it has been argued that the Lyth bound requires $r \ll 1$ in models of
inflation based on well-motivated particle physics.  In Section 3, we
will comment on these points and discuss the implications of the Lyth
bound for inflationary model building and for future CMB experiments
designed to detect tensor modes.

\section{The Lyth Bound and the Inflationary Flow Equations}

As mentioned in the Introduction, inflationary models can be described
by an infinite hierarchy of slow roll parameters
\begin{equation}
 $\;$^\ell \lambda_H  = \left ( {m^2_{\rm pl} \over 4 \pi} \right )^\ell 
{ (H^\prime)^{\ell -1} \over H^\ell} {d^{(\ell+1)} H \over d \phi^{(\ell + 1)}}
\label{F1}
\end{equation}
that satisfy a set of `inflationary flow' equations. The inflationary
flow approach has been used by a large number of authors and we refer
the reader to P03 and Kinney \etal (2004) for a summary of the
approach and for the complete set of flow equations.

The models generated here followed the prescription given in Kinney
\etal (2004), except that the flow hierarchy was truncated at
$\ell=10$ (rather than $\ell=5$) and that successful inflationary
models were required to expand by at least $N=55$ e-folds, consistent
with the analysis of Liddle and Leach (2003). Inflation was deemed to
end if $\epsilon>1$, in which case the models were evolved backwards
by $55$ e-foldings to compute various observables, such as the
tensor-scalar ratio $r$, scalar spectral index $n_s$, and the run in
the spectral index $dn_s/d{\rm ln k}$ (using expressions accurate to
second order in slow-roll parameters\footnote{This is a good
approximation if the inflationary potential is smooth, but may be
inaccurate for potentials with sharp features, see Wang, Mukhanov and
Steinhardt (1997).}). For models in which
inflation occurred at the minimum of a potential ($\epsilon \approx
0$), inflation was simply abruptly terminated and observables
calculated using the slow-roll parameters at the end of
inflation. Physically, such cases can be considered as examples of
hybrid inflation (see Liddle and Lyth 2000) when inflation ends
abruptly at some critical field value $\phi_c$, or as examples of
brane-inflation when inflation ends at a critical inter-brane
separation at which open string modes become tachyonic (see Quevedo,
2002, for a review). In practice, any models which achieved $N > 200$
e-foldings were grouped into this category.

We evolved $2 \times 10^6$ models using the above prescription. Figure
1 shows the parameters $r$, $n_s$ and $dn_s /d{\rm ln k}$ for a sample of
these models plotted against each other. The distributions in these
diagrams agree well with the results of previous authors and are straightforward
to understand physically. More than 90\% of the models are of the
`hybrid-type' for which $r$ is negligible (equation 4) and the scalar spectral
index is $n_s > 1$ (since the inflaton is trapped in the minimum of a 
potential). Of the remaining models, those colour coded in blue
in Figure 1 have $r < 0.36$ and so are consistent with current constraints
on the tensor-scalar ratio. The middle panel of Figure 1 shows that these
models span a wide range of $n_s$ and $dn_s /d {\rm ln k}$.

\begin{figure}

\vskip 3.5 truein

\includegraphics{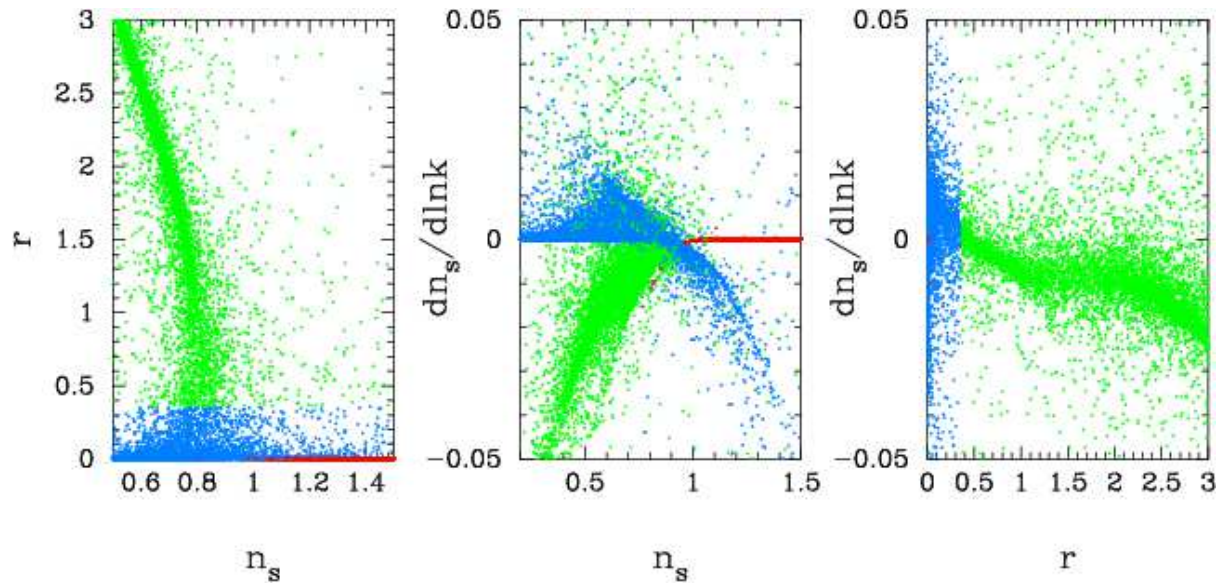}

\caption
{The tensor-scalar ratio, $r$, scalar spectral index, $n_s$, and run in
the spectral index, $dn_s/d{\rm ln} k$, plotted against each other for 
realizations of the inflationary flow equations  described
in the text. The colour coding is as follows: the majority of models
(colour coded red) inflate forever with $\epsilon = 0$ and have
negligible tensor component. In the remaining models,
inflation ends when $\epsilon = 1$ after at least 55 e-folds of inflation.
These models have been colour coded blue if $r \le 0.36$ (the observational
upper limit derived by Seljak \etal 2004) and green if $r >0.36$. }

\label{figure1}

\end{figure}

Figure 2 shows plots of $\vert \Delta \phi\vert $ over the final $55$
e-foldings of inflation computed from equation (3a) plotted against
the tensor-scalar ratio $r$. The figure to the left shows $\Delta
\phi$ for all models within the designated ranges of the abscissa and
ordinate, colour coded so that hybrid-type models are plotted in red,
with the rest of the models plotted in blue. This Figure shows that
for general inflationary models, there is no well defined Lyth
relation of the form (\ref{I7}). Models can be found, for example, with $r
\simgt 1$ and $ \Delta \phi /m_{\rm pl}$ in the range
$0.1$--$0.2$. Likewise, models can be found with low values of
$r\simlt 10^{-4}$ and $\Delta \phi /m_{\rm pl}$ well in excess of
unity. However, on closer inspection, one finds that models lying in
these regions of the diagram have unusual scalar spectral indices. In
the former case, the models have blue spectra with $n_s > 1.1$, and in
the latter case the models have red spectra with $n_s < 0.8$. It is
therefore interesting to plot $r$ against $\Delta \phi$ only for those
models that satisfy observation constraints on the shape of the scalar
power spectrum.

 The tightest observational constraints on $n_s$ and $dn_s/d{\rm ln
k}$ at present come from combining observations of the CMB anisotropies, with
observations of the matter power spectrum deduced from galaxy
redshift surveys and from Ly$\alpha$ lines in the spectra of high redshift quasars
(see {\it e.g.} Viel \etal 2004, Seljak \etal 2004). The recent
results of Seljak \etal  give approximate $2 \sigma$ ranges of
$$ \left. \begin{array} {l}
 \quad 0.92 < n_s < 1.06 \\
 -0.04 < d n_s/d{\rm ln k} < 0.03
 \end{array} \right \}  \eqno(8)
$$
(see Figure 3 of Seljak \etal 2004).
Figure 2b shows $r$ and $\Delta \phi$ for the subset of models that satisfy the
constraints given in (8). (In fact,  although we have imposed a 
constraint on $d n_s/d{\rm ln k}$, it is the constraint on $n_s$ that has the 
most significant effect in defining the distribution of points in Figure 2b.)
The models plotted in  Figure 2b  now delineate a much tighter relationship
between $r$ and $\Delta \phi$, which can be approximated for $r \simgt 10^{-3}$
by 
\begin{equation}
 {\Delta \phi \over m_{\rm pl}} \approx 6 r^{1/4}. \label{F2}
\end{equation}
This relation is our reformulation of the Lyth bound (6).

\begin{figure}

\vskip 3.5 truein

\includegraphics{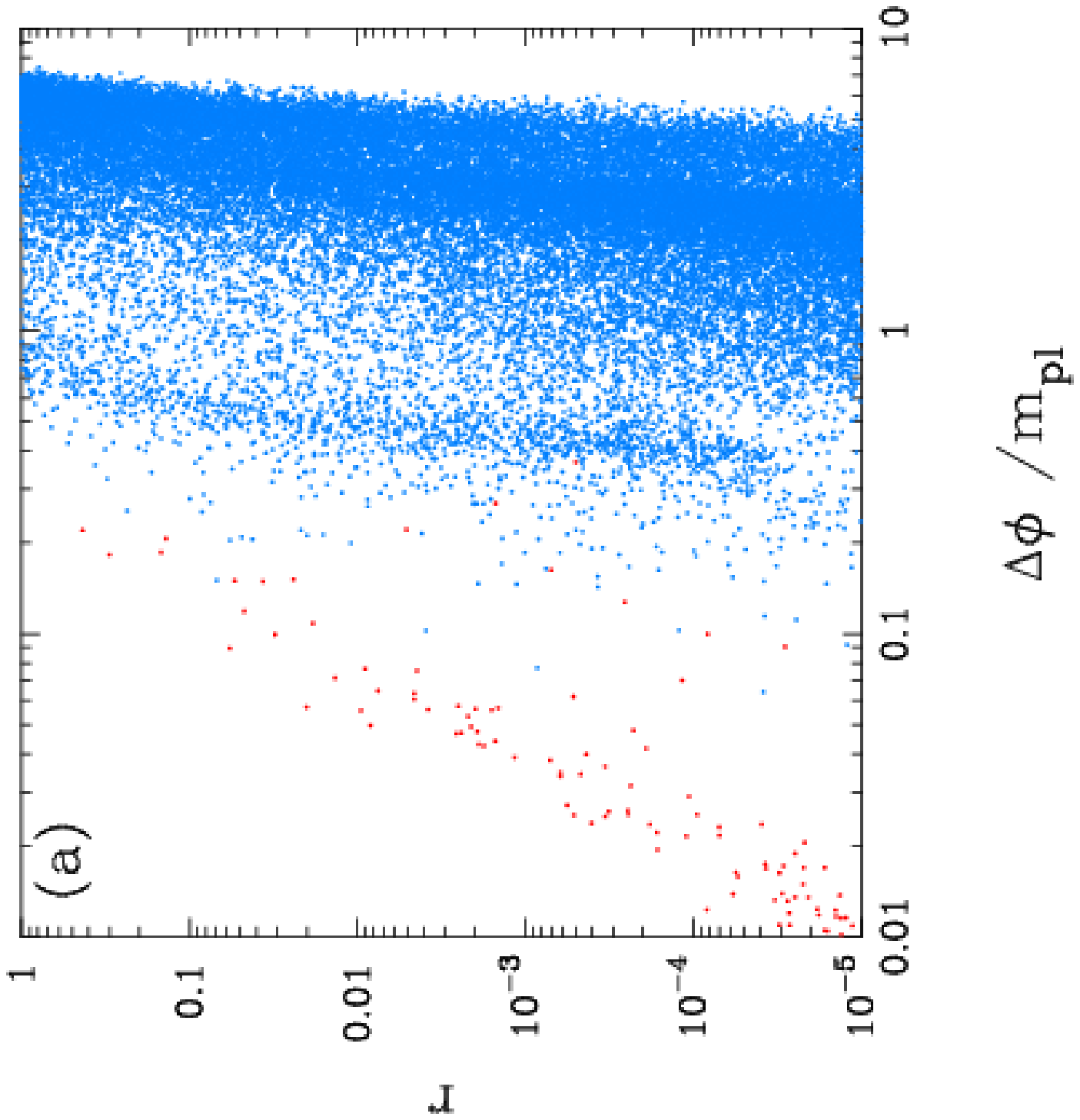}

\includegraphics{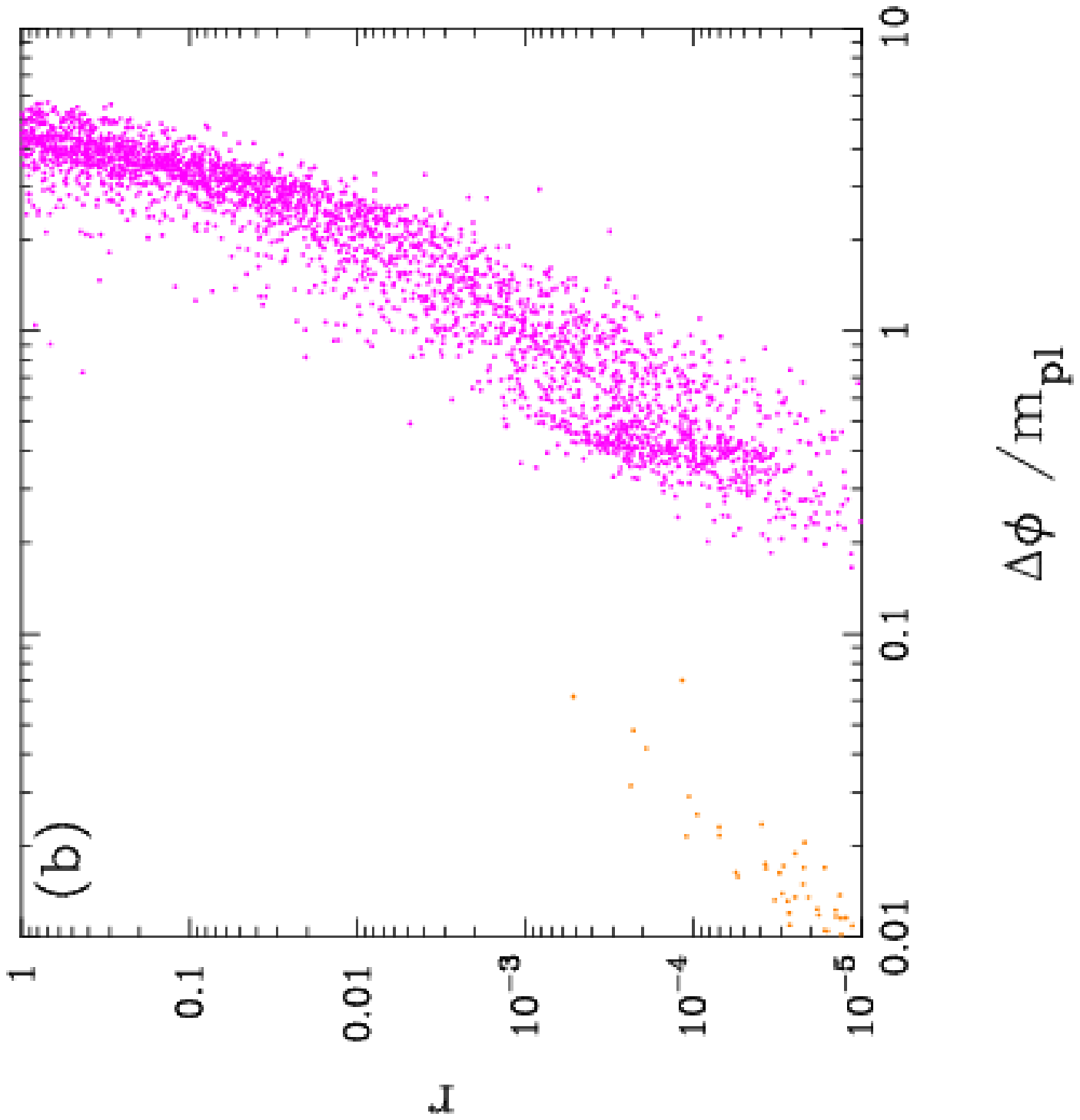}

\caption
{The absolute value of $\Delta \phi$ over the last $55$ e-folds of inflation
plotted against the tensor-scalar ratio, $r$. Figure 2a shows all models
colour coded so that hybrid-type models are shown in red and the rest are shown in
blue. Figure 2b shows the subset of models that satisfy the observational
constraints on  $n_s$ and $dn_s/d{\rm ln} k$ discussed in the text. The colour
coding in this figure is as follows: hybrid-type models are shown in 
orange and the rest of the models are shown in magenta.}

\label{figure2}

\end{figure}

\section{Implications}

 Present observations set the constraint $r < 0.36$ (Seljak \etal
2004). To improve on this limit significantly will require more
sensitive CMB polarization experiments than those done so far. The {\it Planck}
satellite\footnote{http://www.rssd.esa.int/index.php?project=PLANCK}
should be able to achieve a limit of $r \simlt 0.1$ by measuring the
B-mode polarization spectrum, but despite the fact that {\it Planck}
will survey the entire sky from space it is limited in sensitivity by the small
number of polarization sensitive detectors.  Ground based B-mode
optimised experiments, using either bolometer arrays (CLOVER, Taylor \etal
2004), or  large arrays of coherent detectors (QUIET
\footnote{http://astrosun2.astro.cornell.edu/\%7Ehaynes/radiosmm/facs/quiet$\_$i.htm})
are currently being designed that should be able to set limits of $r
\sim 10^{-2}$ in the presence of realistic foregrounds. For $r
\simlt 10^{-2}$, B-mode anisotropies caused by gravitational lensing
of the CMB (Zaldarriaga and Seljak 1998) will need to be subtracted in
order to extract an inflationary tensor component.  However, there are
limits on how accurately a lensing contribution can be subtracted. For
example, even in the absence of instrumental noise and foregrounds,
Lewis \etal (2002) find that a survey of the sky of radius $15^\circ$
leads to a limit of $r \sim 3 \times 10^{-4}$ from the sample variance
of the lens-induced B-modes. This limit could potentially be improved
by a factor of $10$ with a noiseless all sky survey (Kesden \etal
2002, Knox and Song 2002), and perhaps by another factor of $10$ if
the lens-induced B-modes can be mapped accurately enough to
reconstruct the deflection field (Hirata and Seljak 2003). Comparing
with equation (1), under optimistic assumptions the lowest energy
scale of inflation that can be probed by CMB B-mode polarization
measurements is $V^{1/4} \simgt 1 \times 10^{15} \;{\rm GeV}$. More
realistically, the next generation of CMB experiments such as CLOVER and 
QUIET may reach $r
\sim 10^{-2}$,  corresponding to an energy scale of $V^{1/4} \sim 1
\times 10^{16} \;{\rm GeV}$.

Figure 2b shows that to produce a detectable tensor component in any
foreseeable CMB experiment, inflation must necessarily involve large
field variations, $\Delta \phi/m_{\rm pl} \simgt 1 $. In fact, the
dependence of $\Delta \phi$ on $r$ in Figure 2b is so steep that we would
need to achieve $r \simlt 10^{-4}$ to  probe inflationary models with
low field variations of $\Delta \phi /m_{\rm pl} \simlt 0.1$. Thus,
for the foreseeable future, we will only be able to test high-field
models of inflation. Furthermore, it may well prove impossible, given realistic
polarized foregrounds, to probe small-field models of inflation using
the CMB. 

It is often argued ({\it e.g.} Lyth 1997) that the effective potential can be written as a power series
\begin{equation}
V( \phi) = V_0 + \alpha \phi + {m^2 \over 2} \phi^2 + {\beta \over 3} \phi^3
+ {\lambda \over 4 } \phi^4 + \sum_n \lambda_n {\phi^{4 + n} \over m_{\rm pl}^n}, \label{C1}
\end{equation}
with $\lambda_n \sim {\cal O}(1)$, and hence that an effective field
theory description of inflation becomes invalid for field values $\phi
\simgt m_{\rm pl}$. However, as Linde (2004) points out, quantum
gravity corrections to $V(\phi)$ should become large only for $V(\phi)
> m_{\rm pl}^4$ , not $\phi > m_{\rm pl}$ as inferred from equation
(\ref{C1}). This is the rationale behind chaotic inflation models such
as the simple $\lambda \phi^4$ model (which is now excluded
observationally at about the $3 \sigma$ level).  However, it has
proved difficult to find realisations of chaotic inflation motivated
by realistic particle physics. For example, realising the conditions for
slow-roll inflation for either large or small field variations has
been a long standing problem in $N=1$ supergravity.  This may,
however, simply reflect our lack of understanding of supergravity. For example,
it has been pointed out recently that chaotic inflation can be
realised in supergravity models if the potential has a shift symmetry
(Kawasaki \etal 2000, Yamaguchi and Yokoyama 2001), {\it i.e.} the
inflaton potential does not depend on the imaginary part of a complex
field $\Phi$ (see Linde 2004 for a more detailed discussion). It may
therefore be possible to construct particle physics motivated models
of inflation with $\Delta \phi \simgt 1$. Understanding the physics
behind such high-field models is important, for as we have shown in
this paper they are the only class of inflationary models that can be
probed by CMB B-mode polarization experiments in the foreseeable future.

\medskip

\noindent
{\it Acknowledgments:} The research of GPE is supported by the UK Particle
and Astrophysics Research Council.  KM would like to thank the
Institute of Astronomy for their hospitality during a summer visit
when part of this work was done.

\section*{References}
\begin{harvard}

\item[] Bennett C.L., \etal, 2003, {\it Ap J. Suppl. Ser.}, {\bf 148}, 1.

\item[] Easther R., Kinney W.H.,  2003, \PR D, {\bf 67}, 043511.

\item[] Guth A.H., 1981,  \PR D,  {\bf 23}, 347.

\item[] Hirata C.M., Seljak U., 2003, \PR D, {\bf 68}, 083002.

\item[] Hoffman M.B., Turner M.S.,   2001, \PR D, {\bf 64}, 023506.

\item[] Kawasaki M., Yamaguchi M., Yanagida T., 2000, \PR L, {\bf 85}, 3572.

\item[] Kesden M., Cooray A., Kamionkowski M., 2002, \PR L, {\bf 89}, 0113041.

\item[] Kinney W.H.,  2002, \PR D, {\bf 66}, 083508.

\item[] Kinney W.H., 2003, New Astronomy Reviews, 1387.

\item[] Kinney W.H., Kolb E.W., Melchiorri A., Riotto A.,  2004, \PR D, {\bf 69}, 1035161.

\item[] Knox, L., Song, Y-S.,  2002, \PR L, {\bf 89}, 0113031.

\item[] Lewis A., Challinor, A., Turok N., 2002, \PR D, {\bf 65}, 023505.

\item[] Liddle A.R.,  2003 \PR D, {\bf 68}, 103504.

\item[] Liddle A.R., Lyth D.H., 2000, {\it `Cosmological Inflation and Large-Scale Structure},
 Cambridge University Press, Cambridge.

\item[] Liddle A.R., Leach S.M., 2003, \PR D, {\bf 68}, 103503.

\item[] Liddle A.R., Parsons P., Barrow J.D., 1994, \PR D, {\bf 50}, 7222.

\item[] Linde A.D., 1982, {\it Phys. Lett.} B,  {\bf 108}, 389.

\item[] Linde A.D., 1990,  {\it Particle Physics and Inflationary Cosmology},
Hardwood Academic Publishers.

\item[] Linde A.D., 2004, hep-th/0402051.

\item[] Lyth D.H., 1984 Phys Lett B,  {\bf 147},  403L.

\item[] Lyth D.H., 1997 \PRL, {\bf 78}, 1861.

\item[] Lyth D.H., Riotto A.A.,  1999, {\it Physics Reports}, {\bf 314}, 1.

\item[] Peiris H.V., \etal 2003,  {\it Ap. J. Suppl.},  {\bf 148}, 213.

\item[] Quevedo F.,  2002,  {\it Class. Quantum Gravity},  {\bf 19}, 5721.

\item[] Seljak U., \etal, 2004, submitted to \PR D (astro-ph/0407372).

\item[] Taylor, A.C. \etal,  2004, (astro-ph/0407148).

\item[]  Viel M.,  Weller J., Haehnelt M.G., 2004, {\it MNRAS}, {\bf 355}, L23.

\item[]  Wang L., Mukhanov V.F., Steinhardt P.J.,  1997, {\it Phys Lett B}, {\bf 414}, 18.

\item[] Yamaguchi M, Yokoyama J., 2001, \PR D, {\bf 63}, 043506.

\item[]  Zaldarriaga, M., Seljak, U., 1998, \PR D, {\bf 58}, 023003.

\end{harvard}

\end{document}